\documentclass[prl, twocolumn, showpacs]{revtex4}
\usepackage{graphicx, subfigure}
\usepackage{dcolumn}

\def\PRLsection #1 {{\it #1 --  }}  

\begin{document}

\newcommand*{\atow}[0]{$\alpha$$\to$$\omega$}

\title{Impurities block the $\alpha$ to $\omega$ martensitic
  transformation in titanium}

\author{R. G. Hennig$^1$}
\author{D. R. Trinkle$^1$}
\author{J. Bouchet$^2$}
\author{S. G. Srinivasan$^2$}
\author{R. C. Albers$^2$}
\author{J. W. Wilkins$^1$}
\affiliation{$^1$Department of Physics, Ohio State University,
  Columbus, OH 43210\\
  $^2$Los Alamos National Laboratory, Los Alamos, NM 87545}

\date{\today}

\begin{abstract}
  The onset and kinetics of martensitic transformations are controlled
  by impurities trapped during the transformation.  For the \atow\ 
  transformation in Ti, \textit{ab initio} methods yield the changes
  in both the \textit{relative stability} of and \textit{energy
    barrier} between the phases.  Using the recently discovered
  transformation pathway, we study interstitial O, N, C;
  substitutional Al and V; and Ti interstitials and vacancies. The
  resulting microscopic picture explains the observations,
  specifically the suppression of the transformation in A-70 and
  Ti-6Al-4V titanium alloys.
\end{abstract}


\maketitle

Impurities control the onset and kinetics of martensitic phase
transformations---diffusionless structural transformations proceeding
near the speed of sound~\cite{Martensite}.  The abundant use of
martensitic materials in engineering technologies is not underpinned
by a microscopic understanding of the effects of impurities during the
transformation.  The \atow\ martensitic transformation in Ti lowers
its toughness and ductility, affecting its use in the aerospace
industry. The 2 to 9~GPa range in the published transition pressures of
nominally pure Ti~\cite{Jayaraman63,Zilbershtein75} is believed to be
caused by sample purity~\cite{Sikka82,Greeff01}. Systematic studies
show a large sensitivity to small amounts of impurities~\cite{Vohra77,
  Gray92}. Two commercial Ti alloys---A-70 and Ti-6Al-4V---show no
transformation up to 35~GPa~\cite{Gray92}.

Impurities pose two theoretical challenges: The effect on the
\textit{relative stability} of the two phases and the \textit{energy
  barrier} of the transformation.  Understanding both of these
requires the atomistic pathway for \atow\ to model the impurity motion
during the transformation.  The near sonic speed of the transition
prohibits diffusion during the transformation, thus requiring
impurities to move collectively with their local environment.
Recently Trinkle {\it et al.} identified from all possible
transformation pathways the \atow\ mechanism with the lowest energy
barrier of 9~meV/atom. 

This paper exploits this fast transformation that traps the impurities
in their known path to systematically study a range of impurities,
including those in the commercial Ti alloys A-70 and Ti-6Al-4V.  (1)
We study interstitial O, N, C; substitutional Al and V; Ti
interstitials and vacancies. These impurities and defects affect the
transformation by shifting the relative stability of and energy
barrier between the $\alpha$ and $\omega$ phases. (2) Interstitial
impurity effects are governed primarily by their size, while
substitutional impurities affect the transformation by changing the
$d$-electron concentration.  (3) The most important impurities, O in
A-70 Ti and Al in Ti-6Al-4V, more than double the transition barrier
and increase the energy of $\omega$ relative to $\alpha$, explaining
the observed suppression of the \atow\ transition.

\PRLsection{Method} The {\it ab initio} calculations are performed
with VASP~\cite{Kresse93,Kresse96b}, a density functional code using a
plane-wave basis and ultrasoft Vanderbilt type
pseudopotentials~\cite{Vanderbilt90,Kresse94}. The generalized
gradient approximation of Perdew and Wang is used~\cite{Perdew92a}. A
plane-wave kinetic-energy cutoff of 400~eV ensures energy convergence
to 1~meV/atom. The k-point meshes for the different structures are
chosen to guarantee the same accuracy.  To avoid unphysical short
distances between the atoms, it is necessary to include the Ti $3p$
states as valence states in addition to the usual $4s$ and $3d$
states.  For the other elements the pseudopotentials describe the core
states as follows: V [Ar], Al [Ne], O, N and C [He].

Impurity locations and formation energies are determined by
relaxations using $4\times4\times3$ and $3\times3\times4$ supercells
for the $\alpha$ and $\omega$ phases with 96 and 108 atoms,
respectively.  This results in a 1\% impurity concentration.  A
$2\times2\times2$ k-point mesh is used.  The atom positions are
relaxed until the total energy changes by less than $1\,$meV,
corresponding to atomic-level forces $F_\mathrm{max} \leq
20\,$meV/\AA.  Altogether this provides an energy accuracy of
0.5~meV/atom.

NEB calculations with variable cell shape at constant pressure yield
the energy barriers of the martensitic
transformation~\cite{Jonsson98}. The pathway is represented by 18
intermediate images of a 48 Ti atom supercell constructed using
$\sqrt{2}\times 2\sqrt{2}\times 2$ cells of the TAO-1 mechanism with a
$4\times4\times4$ k-point mesh.  This results in an impurity
concentration of 2\%.  Comparing the formation energies for the 48
atom supercell with the larger cells of 96 and 108 atoms provides an
estimate of the finite size error of 0.05~eV.  The images are relaxed
until the energy of each image changes by less than 1~meV,
corresponding to atomic-level forces $F_\mathrm{max} \leq
50\,$meV/\AA\ and stresses $\sigma_\mathrm{max} \leq
20\,$MPa. Altogether this yields an energy barrier accuracy of
1~meV/atom, the same as the finite size error. To compare the effects
of different defects, all energy barriers are reported relative to the
$\alpha$ structure for that defect, and in units of meV per 48 Ti
atoms.

\begin{figure}[tb]
  \includegraphics[width=6.0cm]{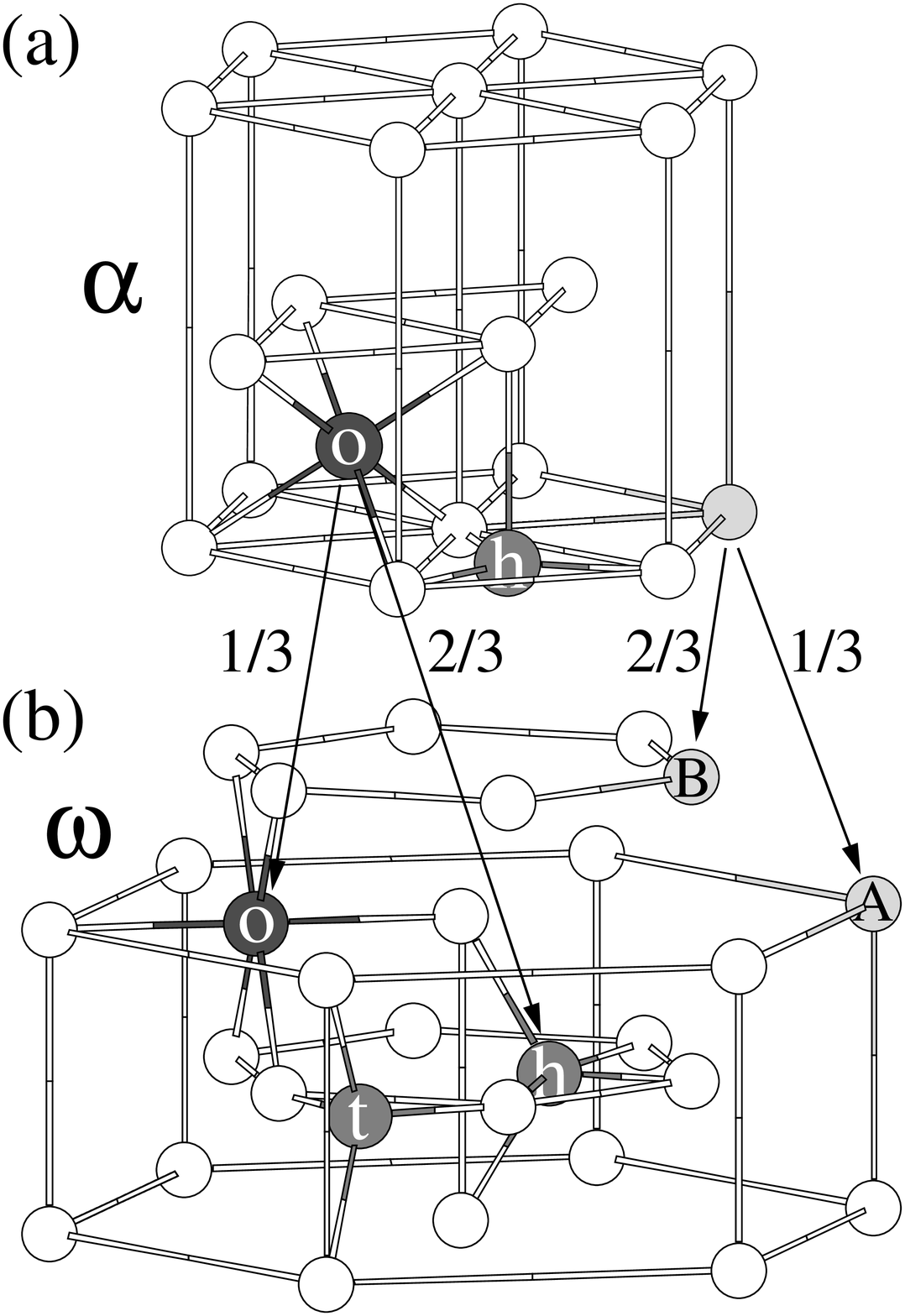}

  \caption {Octahedral (o), tetrahedral (t), and hexahedral (h) sites
    for interstitial impurities as well as $A$ and $B$ sites for
    substitutional impurities in the $\alpha$ and $\omega$ phases. The
    $\alpha$ and $\omega$ phases each contain one unique $o$, $t$, and
    $h$ site. The $\alpha_\mathrm{tet}$ site relaxes to the nearby
    $\alpha_\mathrm{hex}$ site for all three impurities (O, N, C). The
    hexahedral site is a distorted double-tetrahedral site with five
    neighbors.  The arrows indicate the transformation of the impurity
    and lattice sites in the TAO-1 mechanism, and the relative
    proportions.}

  \label{fig:Sites}
\end{figure}

\PRLsection{Interstitial impurities} Figure~\ref{fig:Sites} shows the
location of interstitial impurities and the effect of the
transformation on these sites. The interstitial impurity sites were
found by placing O, N, and C impurities on all interstitial sites in
$\alpha$ and $\omega$ and relaxing the atomic positions.  Each phase
contains one unique octahedral, tetrahedral, and hexahedral site.
However, the tetrahedral site in $\alpha$ is unstable for O, N, and C
and relaxes to the nearby hexahedral site in the basal plane.  The
``hexahedral'' site is a distorted double-tetrahedral site with five
neighbors, formed by a triangle of the basal plane and the two atoms
right above and below the center of the triangle. The instability of
the tetrahedral site is unexpected from crystallographic
considerations~\cite{Conrad81} and is due to atomic relaxations.  In
$\omega$ the tetrahedral and hexahedral sites are both stable, and
they are close in energy and location.

Table~\ref{tab:Impurities}(a) shows the formation energies of the
interstitial defects in $\alpha$ and $\omega$. The energy is measured
relative to molecular O$_2$, N$_2$ and graphite, respectively.  For
both $\alpha$ and $\omega$ the octahedral site is more stable than the
tetrahedral or hexahedral sites.  This is simply related to the size
of the interstitial sites: larger sites have lower formation energies.
Comparison between the $\alpha$ and $\omega$ interstitial impurities
shows that the octahedral formation energy is slightly lower for
$\alpha$; this difference shifts the relative energy of the two
crystal structures, decreasing the stability of $\omega$ over
$\alpha$.

\begin{figure}[tbh]
  \includegraphics[width=8.5cm]{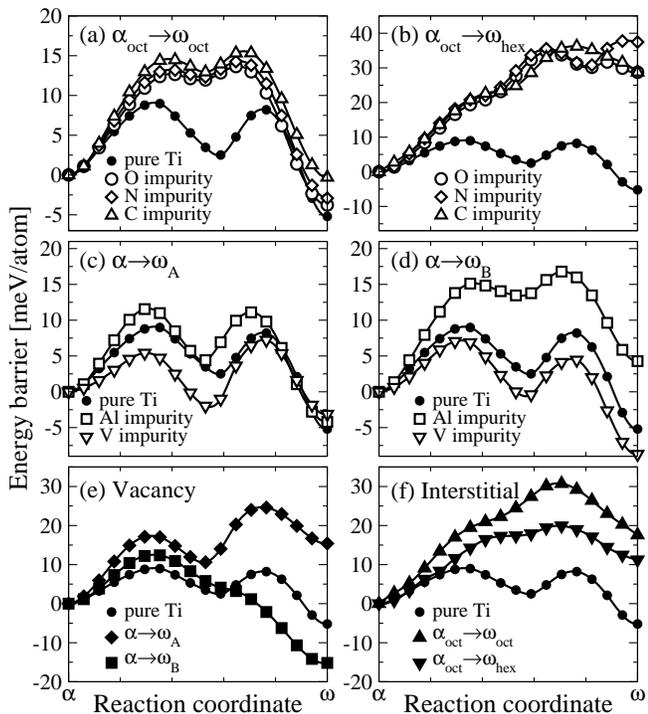}
  \caption {Energy
    barrier for the TAO-1 mappings of interstitial and substitutional
    impurities, as well as vacancies and interstitials in Ti relative
    to the $\alpha$ phase.  The defect concentration is 2\%.  The
    endpoint energies match the formation energies of
    Table~\ref{tab:Impurities} for a defect concentration of 1~at.\%
    within 1~meV/atom, providing an accuracy estimate for the
    barrier.  The interstitial O, N, and C impurities occupy the
    octahedral sites in $\alpha$ and transform either (a) into the
    octahedral site in $\omega$ or (b) into the hexahedral site, with
    a $1:2$ relative ratio.  The substitutional Al and V impurities
    transform in $\omega$ either (c) to the A site or (d) to the B
    site, with a $1:2$ relative ratio.  Figure (e) shows the barrier
    for vacancies and (f) for the Ti-interstitial defects.  The
    vacancy transforms from the $\alpha$ sites to the A and B sites in
    $\omega$ with a ratio of $1:2$.  The interstitial transforms from
    the octahedral site in $\alpha$ to the octahedral and hexahedral
    sites in $\omega$, also with a ratio of $1:2$.  }
  \label{fig:Barriers}
\end{figure}

Figures~\ref{fig:Barriers}(a) and (b) show octahedral sites in
$\alpha$---favored by O, N, and C---transforming into two possible
sites in $\omega$ and increase the energy barrier for the
transformation.  The TAO-1 pathway breaks the hexagonal symmetry and
transforms a third of the octahedral sites into the $\omega$
octahedral site and two thirds into the $\omega$ hexahedral site.  The
presence of impurities increases the energy barrier regardless of the
final site of the impurities; however, this increase is much larger
for the octahedral to hexahedral pathway than for the octahedral to
octahedral one. For all three impurities, despite their different
chemistry, their similar size results in nearly identical barriers.

\begin{table}[tbh]
  \caption{Formation energies and location of interstitial and
  substitutional impurities as well as vacancies and
  self-interstitials in the $\alpha$ ($P6_3/mmc$) and $\omega$
  ($P6/mmm$) phases. The
  formation energies $E_\mathrm{f}$ for O, N, and C are measured
  relative to molecular O$_2$, N$_2$, and graphite, respectively, and
  for Al and V relative to their fcc and bcc phase, respectively. For
  the vacancies and self-interstitials the formation energies are
  measured relative to the corresponding $\alpha$ or $\omega$ phases.  
  All calculations are performed at 1~at.\% defect concentration.
  For each site, the average nearest neighbor distance after
  relaxation, $R_{NN}$, is similar for all impurities; hence, we show
  the range of values.  The coordination number $Z$ at each site is
  also listed.}
  \begin{ruledtabular}
  \begin{tabular}[c]{l l l c c d d d}
    Site &
    \multicolumn{2}{c}{Wyckoff pos.\footnote{See Ref.~\onlinecite{Wyckoff}}} &
    \multicolumn{1}{c}{$R_{NN}$ [\AA]} &
    \multicolumn{1}{c}{$Z$} &
    \multicolumn{3}{c}{$E_\mathrm{f}$ [eV]} \\
    \colrule\\[-0.7em]
    \multicolumn{5}{l}{(a) \emph{Interstitial impurities}} &
    \multicolumn{1}{c}{O} &
    \multicolumn{1}{c}{N} &
    \multicolumn{1}{c}{C} \\[0.2em]
    $\alpha_\mathrm{oct}$  & 2(a) & (0,0,0) & 
       2.06--2.09 & 6 & -6.12 & -5.10 & -1.58 \\
    $\alpha_\mathrm{hex}$  & 2(d) & $(\frac{2}{3},\frac{1}{3},\frac{1}{4})$ &
       1.91--1.95 & 5 & -4.93 & -3.41 & +0.49 \\[1em]
    $\omega_\mathrm{oct}$  & 3(f) & $(\frac{1}{2},0,0)$
       & 2.03--2.17 & 6 & -6.06 & -4.99 & -1.36 \\
    $\omega_\mathrm{hex}$  & 6(k) & $(x,0,\frac{1}{2})$ 
       & 1.92--2.18 & 5 & -4.47 & -3.03 & +0.61 \\ 
    $\omega_\mathrm{tet}$ & 6(m) & $(x,2x,\frac{1}{2})$ & 1.90--2.02 & 4 &
       -4.38 & -2.83 & +0.78 \\
    \colrule\\[-0.7em]
    \multicolumn{5}{l}{(b) \emph{Substitutional impurities}} &
    \multicolumn{3}{l}{\hspace{1.9em} Al \hspace{1.6em} V} \\[0.2em]
    $\alpha$            & 2(c) & $(\frac{1}{3},\frac{2}{3},\frac{1}{4})$ 
                                               & 2.85--2.92 & 12 &
       \multicolumn{3}{l}{\hspace{1.4em}--0.88 \hspace{.5em}+0.51} \\
    $\omega_\mathrm{A}$ & 1(a) & (0,0,0)       & 2.82--3.00 & 14 &
       \multicolumn{3}{l}{\hspace{1.4em}--0.90 \hspace{.5em}+0.60} \\
    $\omega_\mathrm{B}$ & 2(d) & $(\frac{1}{3},\frac{2}{3},\frac{1}{2})$
                                               & 2.60--3.00 & 11 &
       \multicolumn{3}{l}{\hspace{1.4em}--0.47 \hspace{.5em}+0.33} \\      
    \colrule\\[-0.7em]
    \multicolumn{5}{l}{(c) \emph{Vacancy}} \\[0.2em]
    $\alpha$   & 2(c) & $(\frac{1}{3},\frac{2}{3},\frac{1}{4})$
                       & 2.85--2.94 & 12 && +2.03 \\
    $\omega_\mathrm{A}$ & 1(a) & (0,0,0)
                       & 2.83--3.02 & 14 && +2.92 \\
    $\omega_\mathrm{B}$ & 2(d) & $(\frac{1}{3},\frac{2}{3},\frac{1}{2})$
                       & 2.12--3.03 & 11 && +1.57 \\[0.5em]
    \colrule \\[-0.7em]
    \multicolumn{5}{l}{(d) \emph{Self-interstitial}} \\[0.2em]
    $\alpha_\mathrm{oct}$  & 2(a) & (0,0,0) 
                       & 2.47 & 6 && +2.58 \\
    $\alpha_\mathrm{dumb}$  & 2(d) & $(\frac{2}{3},\frac{1}{3},\frac{1}{4})$
                       & 2.05 & 1 && +2.87 \\
    $\omega_\mathrm{oct}$  & 3(f) & $(\frac{1}{2},0,0)$
                       & 2.33--2.60 & 6 && +3.76 \\
    $\omega_\mathrm{hex}$  & 6(k) & $(x,0,\frac{1}{2})$ 
                       & 2.21--2.55 & 5 && +3.49 \\
    $\omega_\mathrm{tet}$ & 6(m) & $(x,2x,\frac{1}{2})$
                       & 2.15--2.51 & 4 && +3.50 \\
  \end{tabular}
  \end{ruledtabular}
  \label{tab:Impurities}
\end{table}

\PRLsection{Substitutional impurities} Table~\ref{tab:Impurities}(b)
shows the formation energies of the substitutional Al and V for
$\alpha$ and the A and B sites in $\omega$.  These energies are
defined relative to fcc Al and bcc V.  Comparison between the $\alpha$
and $\omega$ substitutional impurities shows that, for both Al and V,
the favored $\omega$ site is lower than the $\alpha$ site.  However,
during the transformation, a random mixture of $\omega$ occupations
will be produced in a $1:2$ ratio as shown in
Figure~\ref{fig:Sites}. In that case, the formation energy combination
$\omega_\mathrm{A}/3 + 2\omega_\mathrm{B}/3$ is \textit{lower} than
$\alpha$ for V, and \textit{higher} for Al.  This is expected to
\textit{increase} the stability of $\omega$ relative to $\alpha$ with
V, and \textit{decrease} the stability of $\omega$ with Al.

Figure~\ref{fig:Barriers}(c) and (d) show how the $\alpha$
substitutional sites for Al and V transform to the two possible
$\omega$ sites and effect the transformation barrier.  The TAO-1
pathway lowers the symmetry and transforms a third of the $\alpha$
sites into $\omega_\mathrm{A}$ sites and two thirds into
$\omega_\mathrm{B}$ sites.  The presence of impurities changes the
energy barrier in a similar fashion as the formation energies.  Since
the shift in the barrier matches the shift in the relative energies of
$\alpha$ and $\omega$, we conclude it is due to the $d$-electron
concentration.  The $d$-electron effect is also seen in the alloying
behavior: Transition metals with large $d$-electron concentrations
such as V, Mo, Fe, Ta stabilize the $\omega$ phase, while simple
metals and early transition metals such as Al favor
$\alpha$~\cite{Sikka82}.

\PRLsection{Titanium vacancies and interstitials}
Table~\ref{tab:Impurities}(c) and (d) show the formation energies of
the vacancies and Ti-interstitial defects in $\alpha$ and $\omega$,
respectively.  The reference energy is the appropriate Ti crystal
phase.  For the vacancies, the $\omega_\mathrm{B}$ site---which
relaxes the 6-fold B-sublattice rings into 5-fold rings---is most
forgiving.  However, during the transformation, a random mixture of
$\omega$ occupations will be produced in a $1:2$ ratio; in that case,
the formation energy combination $\omega_\mathrm{A}/3 +
2\omega_\mathrm{B}/3$ is nearly identical to the $\alpha$ vacancy.
The interstitials are energetically favorable, as is expected for a
closed-packed metal.  While the $\alpha$-octahedral site has the
lowest formation energy, there is an energetically close defect formed
when the tetrahedral interstitial relaxes and forms a $[0001]$
dumbbell configuration.  All the $\omega$ interstitial sites are
nearly degenerate in energy.  The increased formation energy for
$\omega$ interstitials over $\alpha$ interstitials is expected to
shift the relative energy of the two crystal structures, decreasing
the stability of $\omega$ over $\alpha$.

Figure~\ref{fig:Barriers}(e) and (f) show how vacancies and
Ti-interstitials transform to two possible sites in $\omega$ and
increase the energy barrier for the transformation.  The TAO-1 pathway
acts on vacancies and Ti-interstitials as for substitutional and
interstitial impurities, respectively.  The presence of vacancies
increases the barrier in both cases, though the effect is much more
pronounced when transforming to the $\omega_\mathrm{A}$ vacancy.  The
presence of interstitials increases the energy barrier regardless of
the final site of the interstitial; unlike the impurities, there is
not much difference between the octahedral and hexahedral pathways.
Due to their inherently low concentration the self defects will not
have any significant effect on the transformation.

\PRLsection{Discussion} The $\alpha$-$\omega$ phase transition is
sensitive to small energy changes. At zero pressure and temperature,
the $\omega$ phase is about 4~meV/atom lower in energy than
$\alpha$~\cite{Kutepov03} and the energy barrier from $\alpha$ to
$\omega$ is 9~meV/atom~\cite{Trinkle03}.  It takes 10~GPa to overcome
the energy barrier in shock experiments~\cite{Gray92} or 280~K to
overcome the energy difference~\cite{Rudin}.  Hence, we expect a
significant change in transition temperature and pressure due to the
energy shift caused by impurities.

\begin{table}[tbh]
  \caption{Change of the relative energy between $\alpha$ and $\omega$
    and of the energy barrier by impurities in the A-70 and the Ti-6Al-4V
    alloys. The impurity concentrations are taken from
    Ref.~\onlinecite{Gray92}.}
  \begin{ruledtabular}
  \begin{tabular}[c]{l l d d}
    Alloy & Impurity &
    \multicolumn{1}{c}{~~$\Delta E_{\omega-\alpha}$[meV]~~~} &
    \multicolumn{1}{c}{$\Delta E_b$ [meV]} \\
    \colrule\\[-0.7em]
    A-70:        & O (1.10~at.\%)  & +12 & +10 \\
                 & N (0.08~at.\%)  &  +1 &  +1 \\
                 & C (0.07~at.\%)  &  +1 &  +1 \\
                 & Total           & +14 & +12 \\
    \colrule\\[-0.7em]
    Ti-6Al-4V:~~~& Al (10.7~at.\%) & +29 & +31 \\
                 & V  (3.8~at.\%)  &  -3 &  -3 \\ 
                 & O  (0.5~at.\%)  &  +6 &  +5 \\
                 & Total           & +33 & +33 \\
  \end{tabular}
  \end{ruledtabular}
  \label{tab:Energies}
\end{table}

Table~\ref{tab:Energies} shows the effect of impurities in A-70 Ti and
Ti-6Al-4V, especially the significant effect of O and Al.  Overall, we
find the presence of impurities in both alloys suppresses the \atow\ 
martensitic phase transition by increasing both the $\omega$ energy
relative to $\alpha$ and the energy barrier.  The increased energy
barrier is expected to reduce nucleation and to slow down the
transformation.  The change of the $\alpha$-$\omega$ energy difference
shifts the $\alpha$-$\omega$ phase boundary~\cite{Rudin} downward by
several hundred degrees and thus at room temperature increases the
transformation pressure dramatically.

The other interstitial impurities, N and C, have an effect similar to
O.  Despite their different chemistry, all increase the energy barrier
and the relative energy by nearly the same amount.  This indicates the
primary effect is steric, and hence, we expect other interstititals of
comparable size to have similar effects.  In the A-70 Ti and Ti-6Al-4V
alloys the effect of N and C is smaller proportional to their lower
concentration.

In contrast, the substitutional Al and V impurities have opposite
effects due to the change in $d$-electron concentration. As expected
from the alloying behavior of Ti~\cite{Sikka82}, V decreases the
energy of $\omega$ and Al increases it.  We observe the same behavior
for the energy barriers.  The Al impurity reduces the $d$-electron
concentration by two, while V increases it by one; thus Al has a
larger effect on the \atow\ transformation.  Commercial Ti-6Al-4V
alloy contains both; however, the concentration of Al (11~at.\%) is
three times higher than the concentration of V (4~at.\%), explaining
the observed suppression of the \atow\ transformation.

\PRLsection{Conclusion} We determined the energy and location of point
defects in $\alpha$ and $\omega$ Ti and showed how they suppress the
martensitic transformation.  Interstitial O, N, and C occupy the
octahedral interstitial sites in $\alpha$ and $\omega$ and transform
into the octahedral and hexahedral interstitial sites in $\omega$. The
interstitial impurities retard the transformation by increasing the
transformation energy barrier while the substitutional impurities Al
and V influence the barrier by changing the $d$-electron
concentration. The effect of impurities on relative energies and
energy barriers is central to understanding martensitic
transformations.

\begin{acknowledgments}
  This work was supported by NSF and DOE. Computational resources were
  provided by the Ohio Supercomputing Center and NERSC.  We thank
  G.~T.~Gray for helpful discussions.
\end{acknowledgments}


\end{document}